\documentclass[conference]{IEEEtran}
\IEEEoverridecommandlockouts
\usepackage{cite}
\usepackage{amsmath,amssymb,amsfonts}
\usepackage{algorithmic}
\usepackage{graphicx}
\usepackage{textcomp}
\usepackage{xcolor}
\newcommand{\etal}{\textit{et al. }}

\def\BibTeX{{\rm B\kern-.05em{\sc i\kern-.025em b}\kern-.08em T\kern-.1667em\lower.7ex\hbox{E}\kern-.125emX}}

\pdfoutput=1
\usepackage{caption}
\usepackage{amsmath}
\usepackage{amssymb}
\usepackage{graphicx}
\usepackage{enumitem}
\usepackage{algorithm2e}
\newcommand{\removelatexerror}{\let\@latex@error\@gobble}
\usepackage{multirow}
\IEEEsettopmargin{t}{0.80in}
\usepackage{url}

\begin{document}

\title{SmartBullets: A Cloud-Assisted Bullet Screen Filter based on Deep Learning}

\author{
    \IEEEauthorblockN{Haoran Niu, Jiangnan Li, Yu Zhao}
    \IEEEauthorblockA{Department of Electrical Engineering and Computer Science
    \\University of Tennessee, Knoxville
    \\\{hniu1, jli103, yzhao64\}@vols.utk.edu}
}
\maketitle

\begin{abstract}

Bullet-screen is a technique that enables the website users to send real-time comment `bullet' cross the screen. Compared with the traditional review of a video, bullet-screen provides new features of feeling expression to video watching and more iterations between video viewers. However, since all the comments from the viewers are shown on the screen publicly and simultaneously, some low-quality bullets will reduce the watching enjoyment of the users. Although the bullet-screen video websites have provided filter functions based on regular expression, bad bullets can still easily pass the filter through making a small modification. 

In this paper, we present SmartBullets, a user-centered bullet-screen filter based on deep learning techniques. A convolutional neural network is trained as the classifier to determine whether a bullet need to be removed according to its quality. Moreover, to increase the scalability of the filter, we employ a cloud-assisted framework by developing a backend cloud server and a front-end browser extension. The evaluation of 40 volunteers shows that SmartBullets can effectively remove the low-quality bullets and improve the overall watching experience of viewers.

\end{abstract}


\section{Introduction}

Bullet-screen, also known as danmaku or DanMu, allows the viewer to send real-time comments, called bullets, that publicly fly across the screen when watching a video. As shown in Fig. 1, the bullets will overlay the video screen directly and will be publicly viewable to all users who watch the video. 

In recent years, danmaku-enabled videos rapidly become popular, especially in East Asian countries, like China and Japan. Bilibili, one of the famous Chinese bullet-screen video websites, ranked 48th in all integrated websites all over the world in April 2019, according to Alexa \cite{alexa}, a well-known web traffic analysis company. Different from traditional video comments and reviews which are static and only allow users to remark the whole video, bullet-screen systems enable a user to create and view comments of a specific scene in a video, which provides users a more direct way to interact with each other and creates a real-time emotional sharing experience. In the rest of the paper, we will use bullet-screen and danmaku interchangeably for simplicity.

\begin{figure}[htbp]
\centerline{\includegraphics[width=0.85\linewidth]{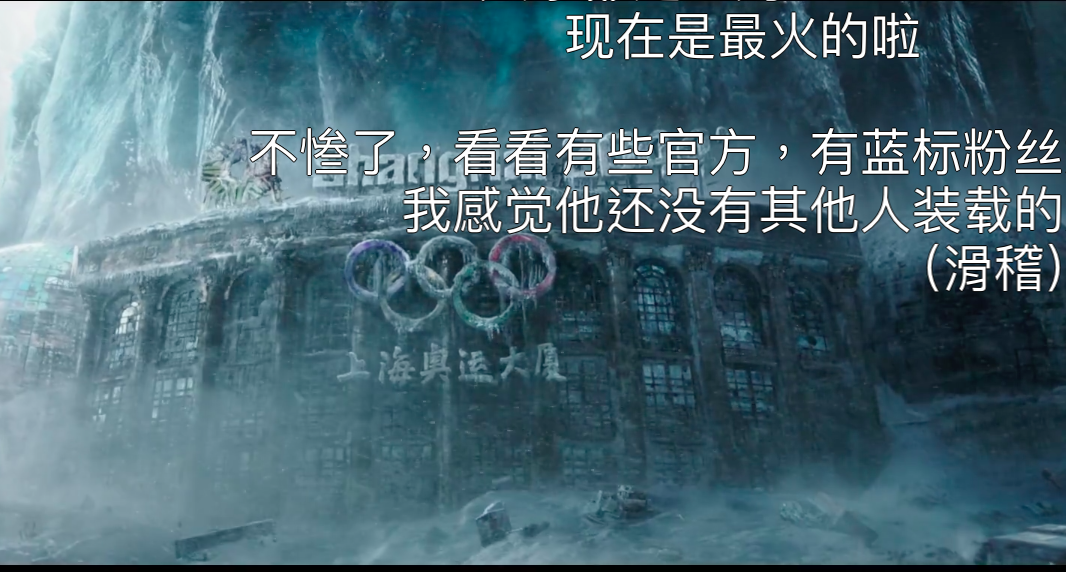}}
\caption{Danmaku Example from Bilibili.com. (Comment from top to bottom: User A: This is the most popular one right now; User B: Looks at the official site; User C: I feel there are others; User D: Smirk)}
\label{fig}
\end{figure}

Generally, a video's bullets are publicly viewable to all the video watchers in danmaku system. Explanatory and humorous comments will strike a chord with the viewers, and further, enhance the interactions between the viewers. On the contrary, there is a risk that low-quality bullets, such as rude and aggressive comments, may cause discomfort to users and reduce the overall watching enjoyment. In order to solve this, famous bullet-screen video websites like Bilibili and Tencent Video have provided basic bullet filter functions to remove bullets according to user's setting. Undesired bullets will be removed according to the position in the screen, font, size, and keywords blacklist. However, due to the diversity and flexibility of natural language, regular expression based bullet filter may have a high false negative rate and can be easily bypassed by making small modifications to the original comment. Therefore, there is a need for a more reliable bullet filter that can sieve bullet comments in accordance with bullet content intelligently to reduce the website maintenance cost and provide a more friendly comment sharing environment to the community.

\begingroup
\begin{figure*}[htp]
\begin{center}
\centerline{\includegraphics[width=0.85\linewidth]{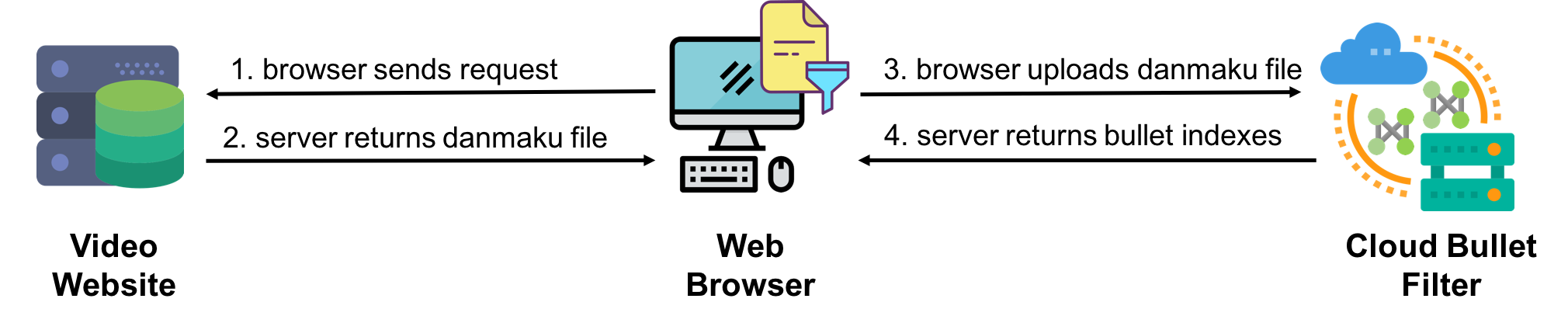}}
\label{fig}
\end{center}
	\setlength{\abovecaptionskip}{-3pt}
	\setlength{\belowcaptionskip}{-10pt}
\caption{SmartBullets Framework}
\label{basic-construction}
\end{figure*}
\endgroup

Natural language processing (\textbf{NLP}) is sub-field artificial intelligence technique that allows the computer to automatically analyze, understand and represent human language \cite{hirschberg2015advances}. In recent years, deep learning technologies, such as convolutional neural network (\textbf{CNN}), recurrent neural network (\textbf{RNN}) and long short-term memory (\textbf{LSTM}), have produced state-of-the-art results in natural language processing field \cite{8416973}. Deep learning-assisted NLP models have been widely employed in many practical applications, such as POS tagging \cite{huang2015bidirectional}, sentiment classification \cite{kim2014convolutional}, and machine translation \cite{gehring2017convolutional}. \cite{fang2015sentiment} uses NLP to perform sentiment analysis of online product review data on Amazon. \cite{liu2016latent} trains a bidirectional RNN to synthesize control program based on natural language input. \cite{dos2014deep} proposes CharSCNN, a deep convolutional neural network to do sentiment analysis of short texts, like Twitter messages. There are also  studies that make use of deep learning to extract information of video from bullet-screen comments \cite{wu2014crowdsourced} \cite{lv2016reading} \cite{chen2017personalized} \cite{he2016predicting} \cite{Chen:2018:FVA:3184558.3186584}. However, to the best of our knowledge, there is no previous study on using deep learning based natural language processing technologies to study the quality of danmaku. 

In this paper, we propose SmartBullets, a cloud-assisted bullet filter framework based on deep learning technologies. SmartBullets consists of a bullet filter that runs on the cloud server and a script program which is embedded into the user's web browser. We utilize a trained CNN model to category bullet comments into two classes according to their qualities. Users can simply enable the filter function through the extension's interface so that bullets with low quality will be filtered. To demonstrate the effectiveness of our framework, we design and implement SmartDanmu, a public Google Chrome extension that can help the user remove the undesired bullets on Bilibili website. We claim that our framework can also be applied on other danmaku-enabled video websites with necessary modification according to the websites' APIs. Our main contributions can be summarized as follow:

\begin{itemize}
\item We review and analyze the research related to bullet-screen comments processing and application, and identify there is a need for more intelligent bullets filtering functions.

\item Based on the state-of-the-art natural language processing techniques, we design and implement a cloud-assisted bullet-screen filtering framework, including a CNN based bullet quality classifier that running on the cloud server and SmartDanmu, a public and convenient front-end Google Chrome browser extension to enhance the scalability.

\item We evaluate our prototype with 40 volunteers. The summary of the survey shows that our bullet filter can effectively remove low-quality bullets and enhance the users' overall enjoyment while watching danmaku videos. We also open-source our code to encourage the research in danmaku community.
\end{itemize}

The rest of paper is organized as follow. Section II gives the introduction of the related research work on Danmaku. Section III introduces the complete design of SmartBullets, including the whole framework, the CNN model for bullet-screen comment classification, and the design of SmartDanmu Chrome browser extension. The detailed description of the framework implementation will be presented in Section IV. After that, Section V introduces the evaluation result. Future work will be discussed in Section VI. Finally, Section VII summarizes the paper.

\section{Related Work}

Inspired by the rapidly increased social media impact of bullet-screen videos, there are more and more studies related to danmaku proposed. \cite{he2018exploring} analyzed the comment distribution of bullets over natural time and discovers the burst patterns of danmaku system. \cite{wu2014crowdsourced} designed a new application that extracts time-sync tags for video shots by automatically exploiting bullet comments of the video. After that, Lv \etal proposed T-DSSM, a temporal deep structured semantic model which can represent bullet-screen comment into semiotic vectors \cite{lv2016reading}. T-DSSM is further used to label highlight shots in videos.  Chen \etal took advantage of the real-time property of bullets and proposed a personalized keyframe recommendation system \cite{chen2017personalized}. In 2016, He \etal made use of danmaku to predict the popularity of a videos \cite{he2016predicting}. On the other hand, Chen \etal employed deep learning model that trained by a bullet-screen comment dataset to predict the attractiveness of fine-grained videos \cite{Chen:2018:FVA:3184558.3186584
}. Other research related to danmaku in recent years can be found in \cite{7979879} \cite{sakaji2016estimation} \cite{ikeda2015classification} \cite {ping2017video}.

\section{Cloud Assisted Bullet Screen Filter Design}

\subsection{Framework Overview}

Our framework mainly comprises two parties, namely a deep learning based bullet classification model that runs on the cloud server and a front-end script program that embedded into the user's browser. As illustrated in Fig. 2, the overall workflow of SmartBullets can be presented in four steps:

\begin{itemize}
\item \textbf{Step 1:} When a user visits the webpage of a danmaku-enabled video, the browser will initial an HTTP request to the video website server.
\item \textbf{Step 2:} If the website server receives the request for a specific video, the server will return the client all the files that the browser needs, including general webpage files, such as HTML, Javascript, CSS, and a danmaku file that contains all bullets information of the video, usually in JSON or XML format.
\item \textbf{Step 3:} If the user enables our script extension in the browser, the browser will forward the processed danmaku file to the cloud server after received the response from the video website.
\item \textbf{Step 4:} After the cloud server received the danmaku file from the user, it will feed the danmaku data to the bullet filer as input after necessary pre-processing. The bullet filter will determine whether a bullet is in low-quality or not. The cloud server will then return the client a list of the indexes of low-quality bullets that should be removed from the danmaku file according to the prediction result of the filter.
\end{itemize}

After receiving the indexes from the server, the script program will then forward the bullets according to the indexes to the danmaku implementation function. Finally, the user is able to watch the video with the processed high-quality bullets.

\subsection{Deep Learning Based Bullet Filter}
\subsubsection{Raw Dataset}

Benefit from the blossoming of NLP research, there are many benchmark public datasets for sentiment analysis, such as movie review dataset by Stanford University \cite{stanforddataset}, tweets dataset by Kaggle \cite{kaggledataset}, and dataset for Chinese natural language processing \cite{chinesedataset}. However, to the best of our knowledge, there is no public, well-organized dataset related to danmaku quality available. Authors of \cite{lv2016reading} and \cite{wu2014crowdsourced} collected their own dataset for analysis. However, since both \cite{lv2016reading} and \cite{wu2014crowdsourced} utilize danmaku data for video tagging and recommendation, their datasets are not appropriate for bullet filtering. 

Fortunately, popular danmaku video websites, such as Bilibili and Tencent Video, provide web APIs for developers and users to gather danmaku data. We employ a web crawler to gather danmaku data from Tencent Video. We choose Tencent Video because the API provides the upcount number of each bullet, which will be an important reference to determine the bullet quality in our framework. The web crawler runs for around 30 minutes and gathered 100 thousand bullets data from around 120 videos, the vast majority of which are in Chinese. The original danmaku data is in JSON format, and the related attributes of each bullet are shown in Table 1.

\begin{table}[htbp]
\caption{Related Attributes of Bullet Data}
\begin{center}
\begin{tabular}{|c|c|}
\hline
\textbf{Attribute} & \textbf{Meaning} \\
\hline
CommentID & the unique identity of a bullet\\
\hline
Content & the text content of a bullet \\
\hline
Upcount & the up count number that a bullet earns \\
\hline
IsFriend & the number of user's friends' upcounts \\
\hline
IsOp & the number of user's opponents' upcounts \\
\hline

\end{tabular}
\label{tab2}
\end{center}
\end{table}

We empirically calculate the overall score $S$ of each bullet using equation (1). The score $S$ is used as a reference to label the records in the next steps.

\begin{equation}
S = Upcount - IsFriend + IsOp\label{eq}
\end{equation}

We clean the raw dataset by removing all type errors and unrelated attributes. We finally format the raw dataset with each record to be $(Content, S)$ pair.
\subsubsection{Pre-processing}

The raw dataset contains too much noise that will destroy the deep learning model. Therefore, basic pre-processing is need before feeding the data to the model. In summary, our pre-processing to the raw dataset mainly includes three steps, as shown below.

\begin{itemize}

\item \textbf{Tokenization: } Since a Chinese sentence is written without space between words, it needs to be split into word segments before the word embedding process. In this step, bullets from the raw dataset are split into various length word lists. 

\item \textbf{Stopwords: } Stopwords are the most common words in language and in general have little contribution to the meaning of the sentence, such as `what'. In this step, stops words in each bullets string will be removed according to a stopwords dictionary.

\item \textbf{Aggregation: } One special property of bullet screen comments is that there are generally a number of repeated bullets in the comment data. In our dataset, two bullets with same content may have different upcount number due to the time of post, which will disturb the overall distribution of the dataset. In this step, we aggregate all repeated bullets by summing up the corresponding scores. 

\end{itemize}

After the above three steps process, each record of the raw dataset will become a unique bullet words list with an overall score, and there are around 30 thousand records left in the dataset. We manually select 11541 low-quality records that we believe are offensive and rude, and we label these records as negative. We randomly pick 11541 records from the rest and label them as positive. Finally, We finish the dataset and are ready for the model training.

\subsubsection{Model Structure}

\begin{figure}[htbp]
\centerline{\includegraphics[width=90mm,scale=1]{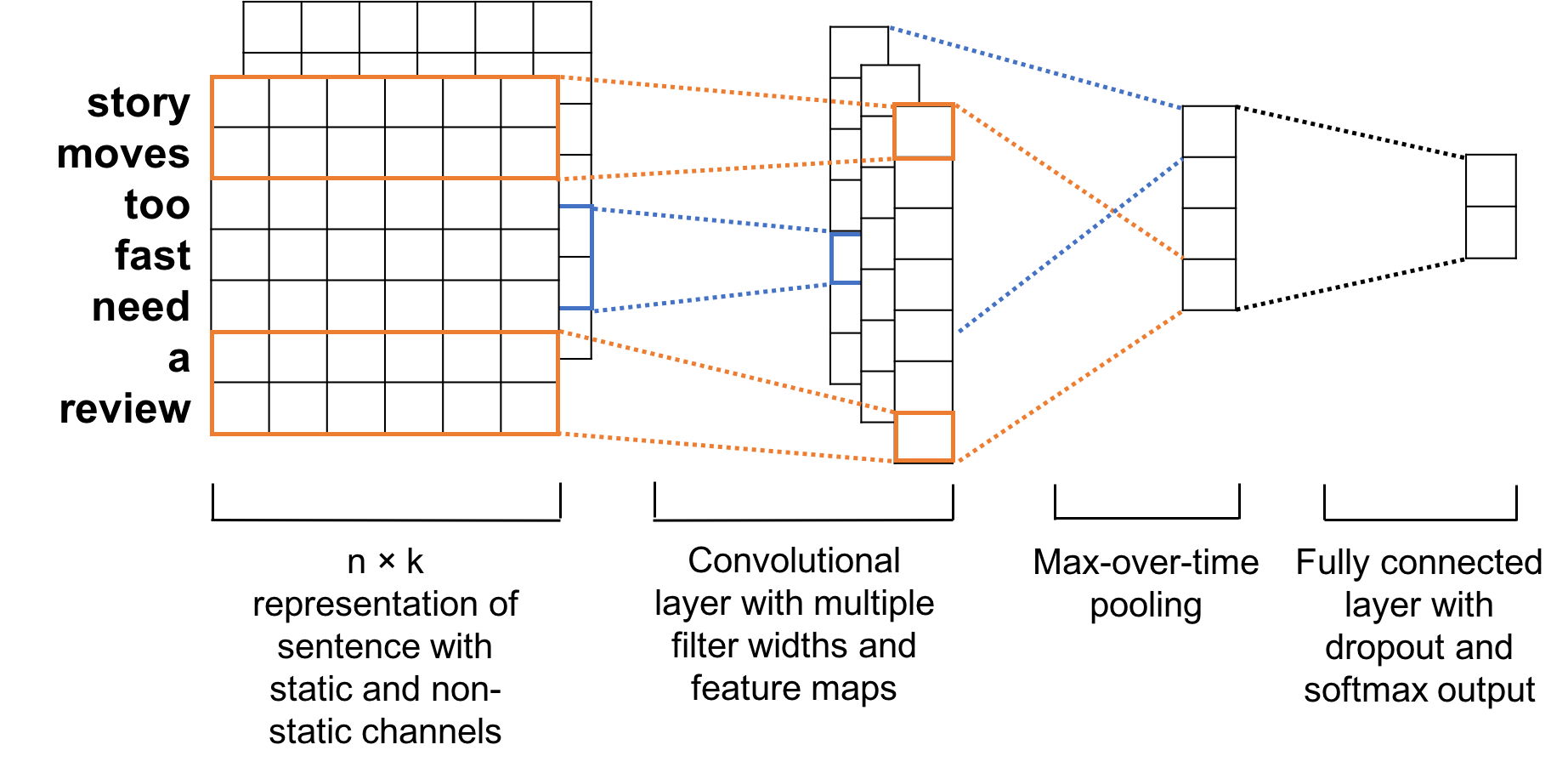}}
\caption{CNN Model Architecture for Sentence Classification from \cite{kim2014convolutional}.}
\label{fig}
\end{figure}

In 2014, Yoon Kim proposed a CNN model trained on top of pre-trained word vectors for sentence-level classification tasks and achieves remarkable experiment result \cite{kim2014convolutional}. As shown in Fig 3, the first part of the model is a hidden word embedding layer to represent sentences, followed by a convolutional layer with multiple filter widths and feature maps. After that, a max pooling layer is employed. Finally, a fully connected layer with dropout is used followed by the softmax output.

We employ the model architecture described in \cite{kim2014convolutional} in our framework with necessary modifications of Chinese word processing to train the bullet classifier. 

\subsection{Front-end Chrome Extension}

\subsubsection{Bullet Screen Implementation Technique}

Generally, a danmaku video website, such as Bilibili and Tencent Video, will have an independent server that maintains the database of the danmaku files and responses the file requests from users. All the bullet comments of each video are stored in a specific format, such as JSON or XML, and includes the attributes of danmaku content and display.  In our project, when a user visits a danmaku video on Bilibili, the user's browser will acquire the webpage of the video which contains a specific query index of the danmaku file, called `cid'. The browser will query the danmaku server with the `cid' and then obtain the danmaku file of the video. The danmaku file will be feed into a JavaScript function that will display danmaku comments that overlay the video according to the danmaku file.

\subsubsection{Design of SmartDanmu Extension}

In order to remove low-quality bullet comments, the SmartDanmu extension is expected to acquire the original danmaku file of the video, upload the related danmaku information to the cloud server, receive the feedback from the server, clean the danmaku file according to the feedback, and input the cleaned danmaku file to the danmaku display function.

Since we employ centered cloud server to process the danmaku filter request from different clients, it is necessary to reduce the overall computation and communication overhead of the server. In our design, the SmartDanmu extension needs to summarize the danmaku file, and only sends a list of ordered danmaku comments to the server. After classifying the bullets, the server will response the extension a list of binary element (0, 1) that follows the order of comments list, where 1 represents positive, and 0 means the corresponding comments are positive and should not be displayed. Finally, the extension will remove the positive bullets and send the cleaned file to the display function.

\section{Implementation}

We implement a prototype of our framework in the laboratory environment, including the CNN-based bullet classifier, a simple cloud web server that handles the requests from users, and the front-end Google Chrome extension that communicates with the cloud server and process the danmaku file. Our source code is available on Github \cite{smartbullets}.

\subsection{Backend}
The bullet classifier and the web server are implemented in Python 2.7. Before training the CNN model, we utilize a popular Chinese words tokenization library Jieba \cite{jieba} to segment the Chinese sentences into words. Jieba is also able to segment English sentences or Chinese sentences with English words. \cite{cnnText} is an open source project in Github that provides a Tensorflow-based implementation of the CNN model in \cite{kim2014convolutional}. We apply the code in \cite{cnnText} to train the bullet classifier with necessary modifications in pre-processing and vocabulary usage. Our model is trained on a server with a 2.40 GHz Intel Xeon CPU and an external NVIDIA GeForce GTX 1080 Ti GPU. We randomly pick 20\% records from the dataset as the testing set and the rest as the training set. We utilize an Adam Optimizer with the learning rate set to 0.001. The training process takes around 30 minutes and can achieve over 93\% accuracy and almost 94\% recall in around 3000 training steps through manually hyperparameter tuning, as shown in Fig. 4.

\begin{figure}[htbp]
\centerline{\includegraphics[width=95mm,scale=1]{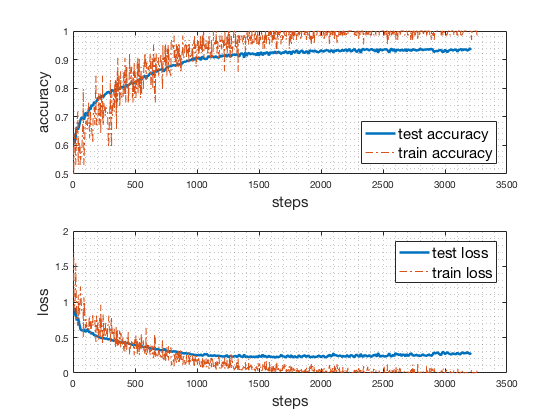}}
\caption{Accuracy and Loss of Model Training}
\label{fig}
\end{figure}

We implement a small Python-based web server to listen to a specific TCP port for the bullet filter requests and use HTTP as the application layer communication protocol between the server and the client. In our experiment, the bullet classifier and the web server are running on a private server in our laboratory with a constant IP address. Our prototype can handle up to 200 filter requests simultaneously.

\subsection{Front-end}

We implement the front-end Google Chrome extension SmartDanmu, which is used to help the user remove low-quality bullets on Bilibili video website. We note that the extension can be further extended to other danmaku enabled websites with essential modification. 

In general, a Chrome extension contains four files, namely a JSON file used for configuration, an HTML file for display, a JavaScript file for functions, and an icon. We firstly modify the JSON file and add the permission of visiting the cloud server. To enable the filter function, we use Ajax Hook \cite{ajaxhook} to block the original Ajax requests of the video webpage and use JavaScript XMLHttpRequest to send a JSON file that contains the bullet comments to the web server through HTTP protocol. Our implementation borrows necessary danmaku-related functions from Pakku \cite{pakku}, an open-source project that merging repeated bullets on Bilibili. The user interface of SmartDanmu extension is illustrated in Fig. 5.

\begin{figure}[htbp]
\centerline{\includegraphics[scale=0.4]{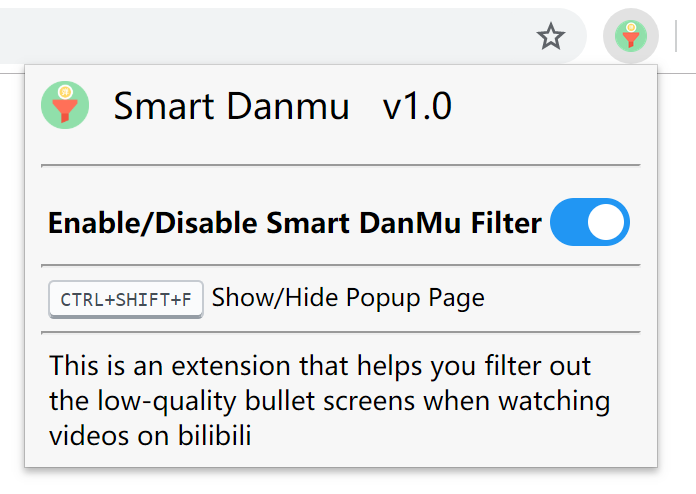}}
\caption{SmartDanmu Chrome Extension User Interface}
\label{fig}
\end{figure}

\section{Evaluation}

We evaluate the prototype of our framework with 40 volunteers, most of whom are college students who are familiar with danmaku control and visit Bilibili video website frequently. Each of the volunteers is asked to watch several top trending videos on Bilibili using Google Chrome browser with and without SmartDanmu extension. After that, each volunteer needs to fill a survey of five statements on the overall user experience, as shown below.

\begin{itemize}
\item \textbf{Legibility: } The instruction of SmartDanmu is clear and easy to read.
\item \textbf{Fluency: } The video webpage is loaded fluently after enabling SmartDanmu.
\item \textbf{Effectiveness: } Most of the low-quality (rude or aggressive) are successfully removed after enabling SmartDanmu.
\item \textbf{User Interface: } The user interface design of SmartDanmu is clear and friendly.
\item \textbf{Experience: } SmartDanmu increase the overall watching experience of danmaku enabled video on Bilibili.
\end{itemize}

For each statement, each volunteer can select one of the three choices, `Agree', `Partially Agree' or `Disagree' to express his/her opinion on that statement. The survey result is shown in Table 2.

\begin{table}[htbp]
\caption{Related Attributes of Bullet Data}
\begin{center}
\begin{tabular}{|c|c|c|c|}
\hline
\textbf{Statement} & \multicolumn{3}{c|}{\textbf{Opinion}} \\
\hline
\textbf{Level} & \textbf{Agree} & \textbf{Partially Agree} & \textbf{Disagree} \\
\hline
Legibility &  37 & 0 & 3\\
\hline
Fluency & 28 & 8 & 4\\
\hline
Effectiveness & 30 & 6 & 4 \\
\hline
UI & 35 & 4 & 1 \\
\hline
Experience & 31 & 7 & 2 \\
\hline

\end{tabular}
\label{tab2}
\end{center}
\end{table}

From Table II, we can learn that 28 out 40 volunteers claim that the webpage loading is still fluent when the SmartDanmu extension is enabled, which means the latency caused by communicating with cloud server is tolerant and even negligible for the video watchers. There are still 12 volunteers declare that they can feel the inconsistency of the video page loading to an extent. We note that a server with more computational and communication resource will improve this delay. Meanwhile, around 75\% participants think SmartDanmu could effectively remove low-quality bullets and increase the overall watching experience compared with the original danmaku videos. 

\section{Future Work}

Our framework is designed to improve the watching experience of users on bullet screen enabled video by filtering out low-quality bullets. However, some users may prefer watching high-quality bullets only. Selecting high-quality bullets is more difficult since they may follow a complex and dynamic distribution over different videos and time, which requires much larger dataset and periodically retraining of the model. Therefore, we suggest that high-quality bullets filtering could be the future work of this paper.

\section{Conclusion}

In this paper, we propose a cloud-assisted user-centered bullet screen filter framework to remove low-quality bullets on bullet-screen enabled video websites. We train a convolutional neural network as the bullet filter that running on the cloud. We also design a front-end browser extension that will communicate with the cloud server and process the danmaku files. We design and implement a prototype in our laboratory and evaluate our framework over 40 volunteers. The survey result shows that our framework is able to effectively remove low-quality bullet screen comments and improve the overall watching experience of users.



\end{document}